\documentclass[nofootinbib,12pt]{revtex4}

\usepackage{amsmath, amssymb,graphicx,epsf,longtable,hyperref}
\def\MR{{\mathbb R}}
\def\MC{{\mathbb C}}

\def\MP{{\mathbb P}}
\def\pro{{\cal P}_l}
\begin{document}
\title{CMB statistical anisotropy, multipole vectors \\and the influence
of the dipole}
\preprint{astro-ph/0603594}
\preprint{IUB-TH-0601}
\author{Robert C. Helling}
\email{helling@atdotde.de}
\author{Peter Schupp}
\email{p.schupp@iu-bremen.de}
\affiliation{International University Bremen\\Campusring 1\\D-28759 Bremen\\
Germany}
\author{Tiberiu Tesileanu}
\email{t.tesileanu@iu-bremen.de}
\affiliation{International University Bremen\\Campusring 1\\D-28759 Bremen\\
Germany}
\begin{abstract}
  A simple algorithm which gives the multipole vectors in terms of the
  roots of a polynomial is given. We find that the reported alignmet
  of the low $l$ multipole vectors can be summarised as an
  anti-alignmet of these with the dipole direction. This
  anti-alignment is not only present in $l=2$ and $3$ but also for
  $l=5$ and higher. This alignment is likely due to non-linearity in
  the data processing. Our results are based on the three year WMAP
  data, we also list corresponding results for the first year data.
\end{abstract}
\maketitle

\section{Introduction}
It is by now a truism that with the availability of high resolution
observations of the microwave sky cosmology has grown from a
somewhat speculative to an empirical science. Especially the WMAP full
sky maps have tightened the error bars on the $\Lambda$CDM
cosmological standard model.

More recently, it has been observed that there are deviations from the
assumption of an isotropic probability distribution of the
fluctuations in the cosmic microwave background
(CMB)\cite{Bernui:2005pz, Copi:2003kt, Schwarz:2004gk, Tegmark:2003ve}
(for a review see \cite{Copi:2005ff}). The source of this anisotropy
is up to now mysterious be it genuinely cosmological, foreground, a
statistical coincidence or a systematic error in the experiment or the
data processing.

In this paper, we contribute to this discussion, although we
are also not able to completely resolve it. To this end, there are two
steps: First, we make contact with the theory of Bloch coherent states. This
leads us to a much simpler algorithm to determine the multipole
vectors: We show that they are simply given as the roots of the
function
\begin{equation*}
  P(z) = \sum_{m=-l}^l \sqrt{2l\choose l+m} a_{lm} z^m,
\end{equation*}
where the $a_{lm}$ are the coefficients of the temperature map expanded in
spherical harmonics.

Secondly, we point out that a more natural interpretation of the
alignment of cross products of multipole vectors is to find the
multipoles themselves in a preferred plane on the sky. To a very good
approximation, this plane is orthogonal to the direction of the dipole
originating from the Doppler shift of the CMB due to the sun's
motion relative to it. 

As, the dipole is not cosmological, it is likely that the
alignment itself is not cosmological either, but due to systematics. We
go on to speculate that non-linearities in the data analysis (due to
feedback from using the dipole to calibrate the detectors) are a
possible cause of such an alignment with the dipole and give 
evidence for this.

During the late stages of this investigation, the WMAP collaboration
has published the data for the first three years of observation. We
have updated our computations and use the new data in the main body of
the paper. For comparison, in Appendix A, we list the multipole
vectors for the first year data, both for the ILC and the Tegmark
et al.\ data sets.  Although, the WMAP team does not find evidence for
non-gaussianity in the low $l$ modes, the multipole vectors and thus
the conclusions of this paper do not change much which is also
anticipated in \cite{Hinshaw:2006ia}.

The structure of this article is as follows: The next section is
mathematical in character and derives our method to compute the
multipoles by introducing the Bloch coherent states. It is followed by
a section where we describe our findings using this method. The next
section presents the Wehrl entropy as a measure of randomness of a
distribution of multipole vectors. Before a
concluding section we offer some hints on the possible origin of the
alignment. In an appendix we give the result of
computations based on the first year WMAP data only for comparison
with the existing literature.

\section{Computing multipole vectors}
The most direct way to represent the cosmic microwave data is as the
temperature as a function of the direction in the sky. Mathematically,
it can be thought of as a real function $T\colon
S^2\to\MR$. Practically, the sky is tessellated into a number of
Healpix\cite{Gorski:2004by} pixels and the LAMBDA\cite{Lambda} archive
provides the temperature data for each of these pixels as obtained
from WMAP observations (let us ignore for the moment the complication due
to contamination from foreground sources which are not cosmological).

For statistical analysis, this pixelised data however is not convenient
and one Fourier transforms (using the {\tt anafast} program) the
temperature data using spherical harmonics
\begin{equation*}
  T(\theta,\phi) = \sum_{l=0}^\infty\sum_{m=-l}^m a_{lm}Y_{lm}(\theta,\phi).
\end{equation*}
We use a normalisation such that 
\begin{equation*}
  \int_{S^2}d\Omega\, Y_{lm}^* Y_{l'm'}=
  \delta_{ll'}\delta_{mm'}.
\end{equation*}
The spherical harmonics form a basis of the vector space of complex
functions on the sphere and when stressing this aspect we denote
$Y_{lm}$ also as $|l,m\rangle$. Note that the subspaces spanned by the
$2l+1$ vectors $|l,m\rangle$ for fixed $l$ form the irreducible
representation of spin $l$ of $SO(3)$, the relevant symmetry group for
the problem.

The temperature is of course a real function. This is reflected in the
fact that 
\begin{equation*}
  a_{l,-m} = (-1)^m a_{lm}^*.
\end{equation*}
This will be important later on.

In an isotropic universe, the $a_{lm}$ for $m\ge 0$ would be
independent random variables of zero mean and a variance which only
depends on $l$ but not on $m$:
\begin{equation*}
  \langle a_{lm}\rangle = 0,\qquad \langle |a_{lm}a_{l'm'}|\rangle =
  c_l\delta_{l,l'}\delta_{mm'}.
\end{equation*}
If the $a_{lm}$ are drawn from a Gaussian distribution, the $c_l$
which are estimated as
\begin{equation*}
  c_l = \frac 1{2l+1}\sum_{m=-l}^l |a_{lm}|^2
\end{equation*}
contain all information about the CMB. In fact, the WMAP measurement
of the $c_l$ has given spectacular credibility to the $\Lambda CDM$ model, at
least for $l>10$. Judging from \cite{Hinshaw:2003ex,
  Bennett:2003bz, Hivon:2001jp}, the original data analysis pipeline
to convert the time series of detector data to pixel data was
optimised to obtain the maximally likely values of the $c_l$ rather
than the $a_{lm}$.

Isotropy of the CMB (more specifically of the random distribution from
which the $a_{lm}$ are drawn) however is an assumption that needs to be
checked. As the $c_l$ are invariant under $SO(3)$ rotations, the
information about isotropy is independent or ``orthogonal'' to the
$c_l$ in the $a_{lm}$. There is a series of papers including
\cite{Bernui:2005pz, Copi:2003kt, Schwarz:2004gk, Tegmark:2003ve}
(for a review see \cite{Copi:2005ff}) noting that in fact the
assumption of isotropy does not hold, at least for small $l$.

For the analysis of isotropy, however, the decomposition of the temperature in
terms of spherical harmonics is not optimal as these are defined with
respect to a choice of $z$-axis. Thus one uses a representation in
terms of ``mutipole vectors'' (going back to
Gau\ss\cite{Weeks:2004cz}) that does not require a choice of reference
direction but transforms covariantly under rotations.

Let us fix an $l$ and denote by $T_l$ the spin $l$ part of $T$:
\begin{equation*}
  T_l(\theta,\phi) = \sum_{m=-l}^l a_{lm}Y_{lm}(\theta,\phi).
\end{equation*}
Instead of using spherical coordinates $(\theta,\phi)$ we can consider
the celestial sphere as the unit sphere $S^2=\{\vec x\in\MR^3\colon
||\vec x||=1\}$ in 3-space. Now, $T_l$ can be written as
\begin{equation*}
  T_l(\vec x) = d_l\pro \prod_{i=1}^l (\vec v_{li}\cdot\vec x).
\end{equation*}
Here $d_l$ is proportional to $\sqrt{c_l}$ (and we will ignore it from
now on), the $\vec v_{li}$ are unit (``multipole'') vectors and $\pro
= \sum_{m=-l}^l |l,m\rangle\langle l,m|$ is the projector to  the spin
$l$ representation. The latter is needed, as a single $\vec
v_{li}\cdot\vec x$ has spin one and the $l$-fold product is a tensor
product of representations which has contributions in spins less than
$l$ as well. Concretely, $\pro$ subtracts terms like $\vec x^2\, (\vec
v_{l1}\cdot \vec v_{l2})\,(\vec v_{l3}\cdot \vec x)\cdots(\vec
v_{ll}\cdot \vec x)$ to make the tensor $\vec
v_{l1}\otimes\cdots\otimes \vec v_{ll}$ totally trace free. Obviously,
when computing scalar products with functions of spin $l$, we can
leave out the projector $\pro$.

The $2l+1$ real components of the $a_{lm}$ contain the same
information as $d_l$ and the $l$ unit vectors $\vec v_{li}$. Strictly
speaking, the $\vec v_{li}$ are only defined up to a sign and thus
strictly live in $\MR\MP^2$. We shall fix this ambiguity by assuming
the multipole vectors to point in directions in the northern
hemisphere.

Note that as opposed to the representation of $T_l$ in terms of the
$a_{lm}$, the representation in terms of multipole vectors is
inherently real and no extra reality condition is needed. In fact, the
antipodal ambiguity can be seen as a consequence of this reality. 

The existing literature gives several ways to compute the $\vec
v_{li}$ from the $a_{lm}$\cite{Copi:2003kt,Katz:2004nj,Weeks:2004cz}.
Here, we will present a different one which seems to be the most
straight forward. We claim that the $\vec v_{li}$ are just the roots
of the polynomial\cite{Dennis1,Dennis2}\footnote{This prescription was
  given before by C. Dennis\cite{Dennis1,Dennis2} as we learned after
  publishing a first version. There, the usefulness for CMB data
  analysis was also pointed out.}
\begin{equation*}
  P(z) = \sum_{m=-l}^l \sqrt{2l\choose l+m} a_{lm} z^m.
\end{equation*}
Strictly speaking, this is of course not a polynomial, but it is a
polynomial divided by $z^l$. Thus, besides $\infty$, it has the same
zero points as a polynomial and we will be sloppy in our language and
call it a polynomial.  A root $z_i\in\MC\cup \{\infty\}$ corresponds
to a point on $S^2$ via the stereographic projection
\begin{equation*}
  z = e^{i\phi} \cot(\theta/2).
\end{equation*}
(note the we use the convention in which $\theta$ runs from $0$ to
$\pi$ unlike the latitude which runs from $-90^\circ$ to $90^\circ$).
The fact that the multipole vectors are only defined up to a sign is
reflected in the invariance of $P(z)$ under the antipodal map
$z\mapsto -1/z^*$, each multipole vector corresponds to a pair of
antipodal roots of $P(z)$. So there are $l$ multipole vectors as
claimed. If $P(z)$ has fewer roots, the remaining multipole vectors
point to the north or south poles of the sphere. This can be seen
after realising that rescaling $P(z)$ by a constant just corresponds
to a rescaling of $d_l$ and does not affect the roots.

To prove this relation between the multipole vectors and the roots of
$P(z)$ we make contact with Bloch coherent states on the sphere.

In analogy with Glauber coherent states, which are eigenstates of the
annihilation operator of the harmonic oscillator and thus have minimum
uncertainty in $x$ and $p$ operators, Bloch coherent states are
defined on the 2-sphere. They can be written as a rotation $R\in
SO(3)$ applied to the highest weight state $|l,l\rangle$. $R$ is
conveniently parametrised in terms of Euler angles $\alpha$, $\beta$,
and $\gamma$ as $R(\alpha, \beta,\gamma)=\exp(i\alpha L_z)\exp(i\beta
L_y)\exp(i\gamma L_z)$.

As $|l,l\rangle$ is an eigenstate of $L_z$, the $\gamma$-rotation acts
only by a phase. As long as we are not interested in overall scalings,
we can ignore this part and actually take $R$ as a function of
$\alpha$ and $\beta$ only and thus living in the coset
$SO(3)/SO(2)=S^2$ which is again a sphere. One observes the
coordinates $\alpha$ and $\beta$ are just the spherical coordinates
$\phi$ and $\theta$ respectively.

Now, we want to compute scalar products of Bloch coherent states with
functions on the sphere defined in terms of multipole vectors. Let us
start with the case $l=1$. We are interested in 
\begin{equation*}
  f(\alpha,\beta) = \langle 1,1| R^{-1}(\alpha,\beta) |\vec v_1\rangle,
\end{equation*}
where we denoted the function $\vec v_l\cdot \vec x$ as $|\vec
v_l\rangle$. 
Obviously, if $\vec v_1$ is in the $z$-direction, the function is just
a spherical harmonic $|\vec e_z\rangle=\sqrt{4\pi/3} |1,0\rangle$.
As $|1,0\rangle$ is orthogonal to $|1,1\rangle$, the scalar product
$f(\alpha,\beta)$ vanishes if the rotation is the identity or minus the
identity. Similarly, for a general vector $\vec v_1$, the scalar
product vanishes if $R$ rotates the $z$-axis in the direction of $\vec
v_1$. This is the case, if $(\alpha,\beta)$ as spherical coordinates
and $\vec v_1$ denote the same point on the sphere.

For higher $l$, we make use of the fact that $|l,l\rangle$ can be
written as a tensor power of the spin one highest weight: $|l,l\rangle
= |1,1\rangle^{\otimes l}$. Here, as we are taking tensor products of
highest weight states only, there are no contributions at lower
spins. Thus, we can express the scalar product as the product for the
different tensor factors
\begin{equation*}
  \langle l,l|R^{-1}(\alpha,\beta)\pro\bigotimes_{i=1}^l|\vec v_{l,i}\rangle =
  \prod_{i=1}^l\langle 1,1|R^{-1}(\alpha,\beta) |\vec v_{li}\rangle.
\end{equation*}
This product than vanishes if any of its factors vanishes which
happens if the point with spherical coordinates $(\alpha,\beta)$
coincides or is antipodal with any of the vectors $\vec v_{li}$. 

On the other hand, \cite{Lieb,Schupp} explicitly compute 
\begin{equation*}
  \langle l,l|R^{-1}(\alpha,\beta)|l,m\rangle \propto \frac
  1{\sin(\beta)^l}\sqrt{2l\choose l+m}z^m, 
\end{equation*}
with $z=\exp(i\alpha)\cot(\beta/2)$. Applying this to the scalar
product of the Bloch coherent state with the CMB temperature function
we arrive at
\begin{equation*}
  \langle l,l|R^{-1}(\alpha,\beta)|T\rangle \propto \frac
  1{\sin(\beta)^l}\sum_{m=-l}^l \sqrt{2l\choose l+m} a_{lm}z^m.
\end{equation*}
Combine this with the conclusion that this has to vanish if $z$ as a
point on the sphere (via the stereographic projection) coincides or is
antipodal to one of the multipole vectors to identify the multipole
vectors with the roots of the above expression. This concludes our
proof of our method to find the multipole vectors as roots of a
polynomial.

This representation significantly simplifies the evaluation of
multipole vectors as it can be handed to any polynomial root finder
such as {\tt mathematica}. A list of multipole vectors can be
downloaded from {\tt http://mathphys.iu-bremen.de/cmb}

The realisation of multipole vectors as roots of a polynomial also
allows for a simple error analysis: We can just add a small amount of
noise to the $a_{lm}$ and compute how much the multipole vectors move
on the sky. We found that, on average, the position of the multipole
vectors changed by $(0.13+0.04\cdot l) \hbox{deg}/\mu\hbox{K}$ upon
adding random noise of amplitude $1\mu\hbox{K}$.

Another advantage of our way of computing the multipole vectors as
roots of $P(z)$ is that it has a direct generalization to complex
functions on the sphere that the inherently real representation as
products of terms of the form $\vec v\cdot\vec x$ has not. For a
generic complex function, the $a_{lm}$ no longer obey $a_{l,-m} =
(-1)^m a_{lm}^*$. Thus, the roots of $P(z)$ do no longer come in pairs
of antipodal points. Rather the $2l$ roots now are multipole spinors
and no longer just live on the northern hemisphere. This is clear from
the perspective of \cite{Schupp}, where $P(z)$ was constructed from
elementary spinors of spin $1/2$ rather than the multipole vectors
$\vec v\cdot \vec x$ of spin one.

This has a direct application to temperature data in conjunction with
polarisation: We can treat the polarization as the complex phase of
the temperature and work out the multipole spinors for this combined
function. Unfortunately, the current status of the noise in the
polarisation data for low $l$ does not warrant such an analysis, yet,
and we postpone it to future investigation.

\section{Anisotropy in WMAP multipole vectors}
We apply our algorithm to the internal linear combination (ILC) map of
the combined three year WMAP data\cite{Hinshaw:2006ia}. As explained
there, this foreground cleaned map is trustworthy for low $l$. The
existing literature mainly discusses alignment in the quadratic
Doppler shift corrected map of Tegmark et al.\cite{Tegmark:2003ve}
which was based on the first year data. Therefore, in appendix
\ref{olddata}, we also list the multipole vectors for computations
based on that data as well as on the first year ILC map.

\begin{figure}
\includegraphics{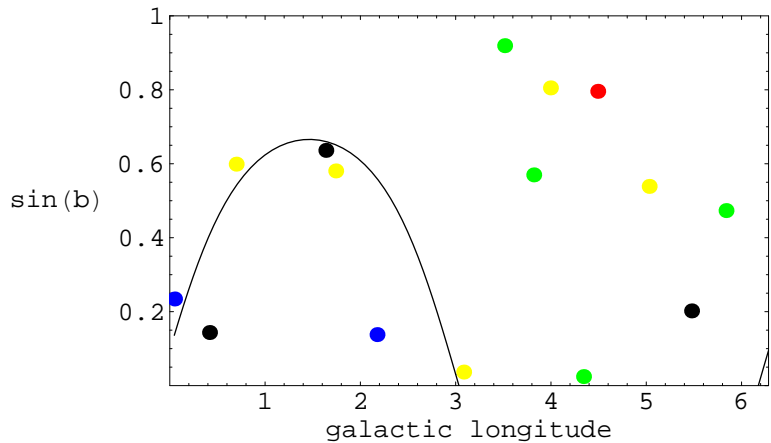}
\caption{\label{multifig}Three year multipole vectors on the
  north galactic hemisphere, the line indicating the plane orthogonal
  to the dipole}
\end{figure}

In table \ref{threeilc} and \ref{threeilc}, we list the multipole
vectors for the three year run WMAP data. Figure \ref{multifig} shows
a graphical representation of the multipole vectors up to $l=5$ on the
northern hemisphere in galactic coordinates (remember that the vectors
are only defined up to reversal).  The colours encode the different
$l$: Red, blue, black, green, yellow stand for $l=1,2,3,4,5$
respectively. Note that by choosing the sine of latitude as a
coordinate we have an area preserving map. This was done to have a
proper impression of probabilities of points falling in a given area
on the celestial sphere.
{\squeezetable

\begin{longtable}{|c||r|r|r|}
\caption{Multipole vectors for three year ILC map. The last
  column lists the angle to the dipole measured such that it is always
  below $90^\circ$.}
\label{threeilc}\\
  \hline
  $l$& latitude & longitude & angle to dipole \\
  \hline\hline
\endfirsthead
\noalign{\begin{center}{\bfseries \tablename\ \thetable{} -- continued
      from previous page}\end{center}}
  \hline
  $l$& latitude & longitude & angle to dipole \\
  \hline\hline
\endhead
\hline
\endfoot
\hline
\endlastfoot
1 &-96.15&48.25&0\\
  \hline
2 & 2.86 & 13.55 & \bf 85.79\\
  & 124.60 & 7.93 & 66.63\\
\hline
3 & 93.83 & 39.51 & \bf 88.21\\
  & 23.89 & 8.26 & \bf 77.13\\
  & -46.28 & 11.66 & 55.17\\
\hline
4 & -25.56 & 28.25 & 56.76\\
  & -111.24 & 1.40 & 48.63\\
  & -158.68 & 66.83 & 36.22\\
  & -141.18 & 34.75 & 35.72\\
\hline
5 & 176.69 & 2.10 & \bf 86.54\\
  & 39.92 & 36.79 & \bf 86.40\\
  & 99.76 & 35.50 & \bf 84.94\\
  & -71.77 & 32.60 & 24.09\\
  & -131.13 & 53.63 & 22.44\\
\hline
6 & 76.41 & 36.95 & \bf 85.46\\
  & 32.70 & 52.10 & 70.60\\
  & -28.79 & 16.25 & 62.95\\
  & -137.11 & 8.86 & 52.28\\
  & -72.33 & 16.99 & 36.82\\
  & -119.85 & 56.76 & 16.62\\
\hline
7 & 24.71 & 8.40 & \bf 76.77\\
  & 92.99 & 20.92 & 69.65\\
  & 175.65 & 26.30 & 69.55\\
  & 5.45 & 41.08 & 67.09\\
  & -11.98 & 33.74 & 61.93\\
  & -73.11 & 20.44 & 33.41\\
  & -136.07 & 66.75 & 27.49\\
\hline
8 & 117.20 & 40.10 & \bf 86.84\\
  & 12.15 & 4.10 & \bf 81.07\\
  & 26.74 & 40.34 & \bf 78.04\\
  & -27.13 & 26.76 & 56.72\\
  & -65.83 & 3.99 & 51.30\\
  & 73.59 & 85.79 & 45.90\\
  & -95.43 & 19.89 & 28.36\\
  & -139.63 & 53.41 & 27.50\\
\hline
9 & 73.09 & 40.06 & \bf 88.82\\
  & 9.56 & 11.45 & \bf 88.36\\
  & 124.20 & 43.40 & \bf 81.73\\
  & 149.11 & 32.43 & \bf 80.51\\
  & -9.58 & 16.84 & \bf 75.27\\
  & -55.44 & 17.85 & 44.83\\
  & -115.52 & 8.66 & 42.83\\
  & -40.11 & 64.13 & 33.53\\
  & -121.87 & 41.15 & 19.50\\
\hline
10 & 4.24 & 10.22 & \bf 89.19\\
  & 69.62 & 44.86 & \bf 86.06\\
  & 140.57 & 40.40 & \bf 78.15\\
  & -4.67 & 25.14 & 72.46\\
  & -50.22 & 0.12 & 62.31\\
  & -108.53 & 0.13 & 49.30\\
  & 152.02 & 80.87 & 45.79\\
  & -57.26 & 29.19 & 35.28\\
  & -126.99 & 25.61 & 33.06\\
  & -88.29 & 62.83 & 15.21\\
\end{longtable}
}
\begin{figure}
\includegraphics{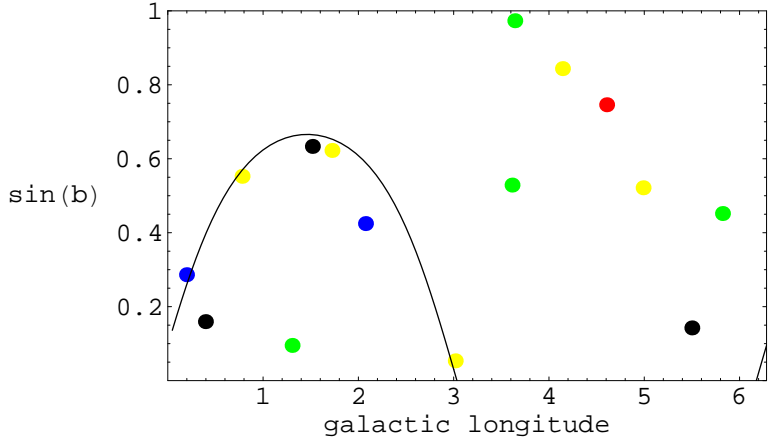}
\caption{\label{year1tdqmap}First year multipole vectors from
  Tegmark et al. map} 
\end{figure}

As noted in the existing literature, there is no obvious clustering of the
multipole vectors at any single point in the sky. There is, however, a
planarity and a more indirect correlation: One can form vector
products of multipole vectors and those are close for $l=2$ and
$l=3$ (see \cite{Tegmark:2003ve,Copi:2003kt}, for a discussion of
dependence on various choices, see \cite{Bielewicz:2005zu}, for the
specifics of the map making procedure \cite{Freeman:2005nx} and for a
review \cite{Copi:2005ff}).
 
More specifically, one computes nor\-ma\-lised vector products like
$
  \vec w_2 = (\vec v_{21}\times \vec v_{22})/\|\vec v_{21}\times
    \vec v_{22}\|
$
and similarly for pairs of $l=3$ multipole vectors. These happen to be
confined to a relatively small region of the sky and this apparent
anisotropy of the random distribution has been termed ``axis of evil''
in \cite{Land:2005ad}. In \cite{Schwarz:2004gk}, the cosine of the
angle between $\vec w_2$ and the three pairs of $l=3$ vectors have
been computed and found to be 0.9531, 0.8719, and 0.8377 and from
these relatively large numbers (note that
$\arccos(0.8377)=33.1^\circ$) it was concluded that the quadrupole and
the octopole are aligned. 

The obvious question is about the origin of this alignment. It could
come from a systematic error in the measurement or post-processing of
the data, it could be a statistical fluke or it could be of genuine
cosmological importance. In this note, we want to argue for the first
possibility.

In order to settle this question, it is worth understanding if
there is another astronomical or cosmological datum in the region in
which the $\vec w$'s are pointing. \cite{Schwarz:2004gk} suggests the
pole of the ecliptic plane, however Bielewicz
et al.\cite{Bielewicz:2005zu} conclude ``While the nominal significance
of these results are confirmed in this paper, we also found that it is
not at all unusual to observe such a strong alignment with one of the
three major axes (ecliptic, galactic or super-galactic), {\em given}
the peculiar internal arrangements of the quadrupole and
octopole. This, it is not the ecliptic correlation {\em per se} that
is anomalous, but rather the quadrupole-octopole alignment.''.

Furthermore, computing the vector products appears to be quite a
derived quantity from the original data. Thus the question about the
geometric significance of the alignment arises. Geometrically, $\vec
w_{2}$ is the vector that is orthogonal to the plane spanned by $\vec
v_{21}$ and $\vec v_{22}$. Thus, an alignment of different vector
products means that all the vectors that cross-multiplied to these
$\vec w$'s lay in a plane on the sky that is orthogonal to the point
on which the $\vec w$'s cluster. So it is rather this plane than the
direction of the vector products that appears to be of significance.

Therefore, it is possible to subsume the different alignments into the
statement that there appears to be a special plane in the sky in which
a lot of multipole vectors happen to be.

What to our mind has not been stressed enough in the previous
literature is that this plane happens to be the plane orthogonal to
the dipole: In Table \ref{multipoles}, we thus present as well the
angle between the multipole vectors and the dipole. There appears to
be an unnaturally large number of multipoles which are nearly
orthogonal to the dipole (highlighted in boldface): Both quadrupole
vectors as well as two of the three octopole vectors and three of the
five $l=5$ multipole happen to be nearly at right angles! (The third
octopole is off by quite a margin even given the alignment claimed in
\cite{Schwarz:2004gk}. However one should realise that the scalar
products of the $\vec w$'s involving this multipole correspond to
angles of $29^\circ$ and $33^\circ$.) For $l=4$, there is
no strong indication of multipole vectors being close to that plane.
For a graphic presentation see\ Figure \ref{multifig}, where the line
indicates the plane orthogonal to the dipole. One can see that not
only multipoles for $l=2$ and $3$ fall on this line, but also three of
the five vectors at $l=5$, a result so far not reported in the
literature\footnote{\cite{Wiaux} applies wavelet methods to the three
year WMAP data and also finds a number of preferred directions in the plane
orthogonal to the dipole direction}.

As far as relevance is concerned, we just mention that the fraction of
a sphere that is at an angle larger than $\delta$ from a given vector
(or its antipodal such that $0\le\delta\le 90^\circ$) is given by
$\cos(\delta)$. Take for example $l=5$, where the probability to find
three vectors at least $\theta=85^\circ$ from the dipole direction is
0.006.

We find the fact that the multipole vectors seem to avoid the
direction of the dipole is a strong indication that the alignment is
not of cosmological origin but rather is an artefact of the data
processing. The dipole (which is roughly a factor of hundred stronger
than the higher spin contributions) is thought to be the first order
Doppler shift contribution from the monopole (again a factor of
thousand stronger) due to the motion of the sun relative to the
rest-frame of the CMB. Thus, any genuine correlation of the CMB with
the dipole could only hold in a rather Ptolemean cosmological
model. In section \ref{systematics}, we therefore discuss possible
systematic origins of such an alignment.

\section{Wehrl entropy}
So far, we have used Bloch coherent states to proove that the roots of
$P(z)$ are in fact the multipole vectors. In \cite{Lieb}, they are
employed to define a notion of entropy for quantum mechanical states
vectors in $SU(2)$ representations. This ``Wehrl entropy''\cite{Wehrl} is a
function on the irreducible representations of spin $l$ is monotonic,
strongly subadditive and positive and thus has all properties of an
entropy. In our notation, it is defined as 
\begin{equation*}
  S_W(T) = -\int_{\raise -0.5ex \rlap{SO(3)/SO(2)}} \,\,d\Omega\phantom{SO(3)}\, \left|
    \langle l,l|R^{-1}(\alpha,\beta)|T\rangle\right|^2\ln\left(\left|
      \langle l,l|R^{-1}(\alpha,\beta)|T\rangle\right|^2\right) 
\end{equation*}
where $\Omega=(\alpha,\beta)$ is a point on the sphere. This entropy
was introduced as a measure of how ``quantum'' a state is and it is a
conjecture due to E. Lieb\cite{Lieb} that it is minimised if the state is
coherent. \cite{Schupp} used the function $P(z)$ to proove this
conjecture for low spin. 
\begin{figure}
\caption{\label{wehrl}The Wehrl entropy for $1\le l\le 20$}
\includegraphics[width=15cm]{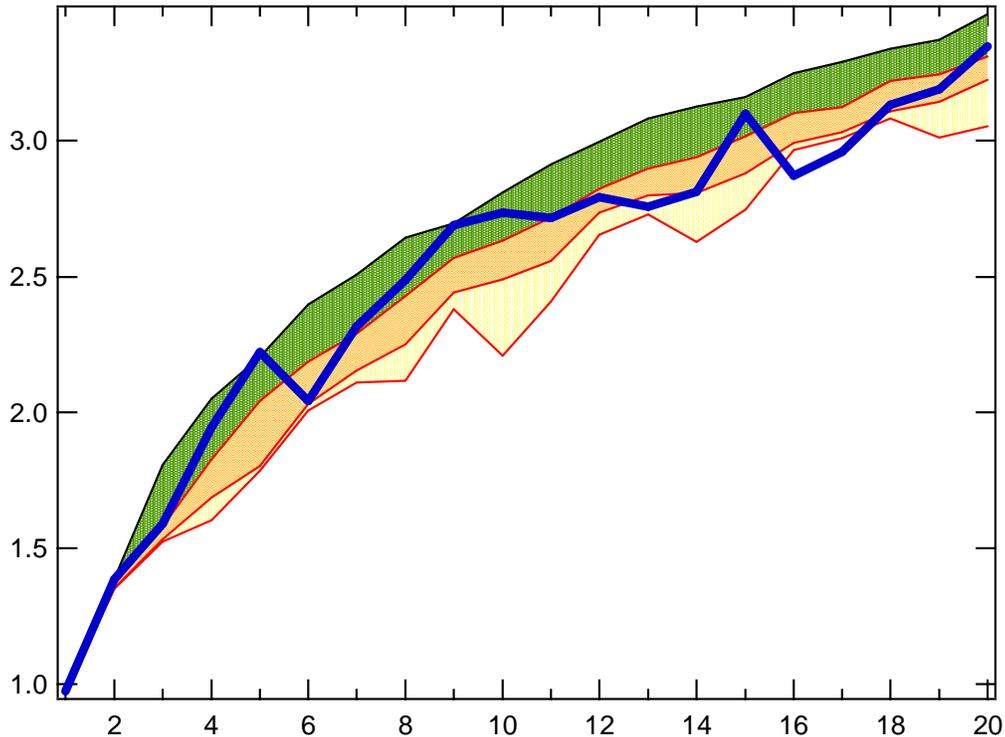}
\end{figure}
Here we ignore the quantum mechanical interpretation and propose the Wehrl
entropy as a measure of ``randomness'' of an ensemble of
multipole vectors. Note, that the coherent states are neccesarily complex
and thus the CMB temperature function $T$ cannot attain the minimal
value assuming the conjecture holds.

We computed the Wehrl entropy for the multipole vectors of the first
year data up to $l=20$. To have a comparison we computed as well the
Wehrl entropy at each $l$ for 50 realisations of random vectors on the
sphere. From these we computed the maximum. Figure
\ref{wehrl} shows the Wehrl entropy for the CMB as measured by WMAP in
blue. The maximum obtained  from random multipole vectors is indicated
by the solid black line. The shaded regions contain 68.3\%, 95.4\%,
and 99.7\% of the simulated collection of random vectors.

\section{\label{systematics}Possible sources of anti-alignment with
  the dipole} 
Finding an apparent correlation between the cosmological multipole
vectors and the dipole arising from the sun's movement calls for an
understanding of an underlying systematic error. In this section, we
want to give some indications on what might be the cause.

Before we come to specifics of the data post-processing, we would like
to point out a peculiarity of the multipole representation: The
ordinary representation of functions on the sphere in terms of
coefficients of spherical harmonics transform in a simple (linear) way
under addition of functions. In contrast, the spherical harmonic
coefficients of a product of functions is quite involved and requires
(as representation theoretically it is a tensor product)
Clebsch-Gordan coefficients
\begin{equation*}
  Y_{l_1m_1}(\theta,\phi)Y_{l_2m_2}(\theta,\phi) =
  \sum_{LM}CG_{l_1m_1,l_2m_2}^{LM} Y_{LM}(\theta,\phi). 
\end{equation*}
The multipole representation, in contrast, is in itself
multiplicative. This implies that the multipoles of a sum of two
functions transform in a complicated way (as the roots of the sum of
two polynomials) and by themselves are not even well defined without
specifying the weights $d_l$. The only thing one can say is that they
will not change too much if a small function is added (see also the
section on noise dependence of the multipole vectors). 

Under products however, the multiple vectors of the product are just
the union of the multipole vectors for the two factors (up to some
complications arising due to the lower order terms arising from the
projector $\pro$). Thus taking products of two functions is the
natural way for multipoles to spread between sectors of different $l$.

We thus consider the following scenario likely for the cause of the
anti-alignment: Instead of recording the proper temperature
distribution in the sky $T(\theta, \phi)$, there is a small
multiplicative error 
\begin{equation*}
  \tilde T(\theta,\phi) = 
\left(1+\epsilon f(\theta,\phi)\right) T(\theta,\phi), 
\end{equation*}
(such an error term was also considered in \cite{Gordon:2005ai}). This
would lead to a mixing between different $l$ modes of $f$ and $T$. The
mode that would be the strongest in the mixing would be the dipole
as the monopole does not mix and the dipole is two orders of magnitude
stronger than the remaining modes.

This would however directly lead to an appearance of the dipole as a
multipole vector for the higher $l$. The anti-alignment would only
appear if there would be a correction for the error $f(\theta,\phi)$
which is slightly off or overcompensating.

The idea, that the dipole cannot be ignored when mode mixing is
considered and indeed gives the main contribution was stressed in
\cite{Vale:2005mt}. However, there the mixing was proposed to be due
to gravitational lensing which was shown in \cite{Cooray:2005my} to be
too weak to explain the alignment.

Another possible source of such multiplicative mode mixing is the
filter used to eliminate galactic foreground
emissions\cite{Hinshaw:2003ex,Hivon:2001jp} using the MASTER
algorithm. Decomposing the filter into spherical harmonics confirms
the expectation that due to rotational invariance, the low $l$ modes
have a large fraction of their power in the $m=0$ modes in galactic
coordinates and that virtually there are only modes with even $l$ due
to invariance of the filter under parity reflections. Our
understanding of the MASTER algorithm is that due to computational
complexity it only gives corrections depending on $l$ and not on $m$
and thus only corrects the $d_l$ or $c_l$ but does not change the
directions of the multipole vectors. Furthermore it was developed for
the BOOMERANG experiment which sees only a small portion of the sky
and thus \cite{Hivon:2001jp} describes Monte Carlo testing only for
$l$ of at least medium size and not the small values of $l$ that show
the alignment.

In our analysis of the filter, we were however not able to show
conclusively that an
application of the galactic filter directly leads to an anti-alignment of
multipole vectors and the dipole.

Besides multiplicative corrections, there is another possibility of
making the dipole influence the higher $l$ modes: The sampling
function could be slightly non-linear:
\begin{equation*}
  \hat T(\theta,\phi) = T(\theta,\phi) + \varepsilon T^2(\theta,\phi).
\end{equation*}
Again, the main contribution would come from the dipole (due to its
strength compared to the other modes), it would even be of a similar
effect if instead of $T^2$ there were a contribution that is
schematically $\hbox{dipole} \cdot T$.

Let us investigate such a contribution. The multipole representation
is covariant under rotations. So it is simplest to go to a frame in
which the dipole points in the $z$-direction. Let us study the
polynomial $P(z)$ in this frame.

Any complex polynomial can be factored into linear factors $(z-a_i)$
in terms of the roots $a_i$. In our case, as noted above, from the
reality of $T^*=T$ we have invariance of $P$ under the antipodal map
$P(z)=P(-1/z)^*$. Thus, the roots always come in pairs $a_i$ and
$-1/a_i^*$. Such a pair contributes a quadratic factor (conveniently
normalised)
\begin{equation*}
  P_a(z) = \frac{a^*}z (z-a)(z+1/a^*) = a^*z +(|a|^2-1)-a/z
\end{equation*}
and the total polynomial $P(z)$ is the product of such factors.

The dipole itself has its roots at the poles, $0$ and $\infty$, thus it comes
with the polynomial 
\begin{equation*}
  P_0(z) = 1.
\end{equation*}
This polynomial obviously lacks the highest and lowest power of $z$.

Similarly, multipoles on the equator (in this frame, orthogonal to the
dipole in a general frame) have $|a|=1$ and thus the constant term vanishes:
\begin{equation*}
  P_a(z) = z/a-a/z\qquad\hbox{for $|a|=1$.}
\end{equation*}

In general, if we treat the roots of a degree $N$ polynomial as
independent random variables, the coefficient of the $z^k$ term is
expected to be of size $\sqrt{N\choose k}$ relative to the other terms
(just from counting the number of terms contributing to $z^k$ when expanding
the product of linear factors). The general form of $P(z)$ above
reflects this, yielding an isotropic distribution of multipole vectors
for independent $a_{lm}$. A polynomial with roots at special locations
(the poles of the sphere, or the unit circle) will thus deviate from
the general form. For example, if all roots lay on the unit circle
(all are orthogonal to the dipole in the frame independent language),
every other coefficient is missing, as for $|a|=1$ the polynomial in
$z$ is effectively in $z^2$.

Let us now see how a small quadratic contamination influences the
location of the multipoles. Consider adding to $P(z)$ at $l$ a small
contribution of $T^2\approx \hbox{dipole} \cdot T$, the result will be
the $P_{l-1}(z)$ for $l-1$ times $P_0(z)=1$. As a result, there will
be a change in the coefficients of $z^{\pm l}$ relative to the
intermediate powers of $z$. We found numerically that such a shift is
well able to move the multipoles towards the equatorial plane.

What could be the origin of such a non-linear contribution? A possible
source of feedback of the dipole into the higher modes comes from the
real time calibration of WMAP on the Earth velocity modulation of the
dipole as described in section 3 of \cite{Bennett:2003bz}. This
procedure uses {\em a priori} knowledge of the dipole from COBE to
compare with the raw WMAP data (without correcting for the galactic
cut). Given the relative strength of the dipole even a comparably
small error in this prior could couple the the dipole into the low $l$
spectrum via non-linear feedback. For example, it could be possible
that minimisation of the dipole could also decrease the $a_{lm}$ for
$|m|<l$ for $l>1$ in the frame aligned with the dipole.

Due to our lack of direct information on the details of this calibration
procedure, we have not attempted to test this hypothesis using for
example Monte Carlo data. Still, we consider it a plausible source of
non-linearity which mixes the dipole into the sectors for $2\le l\le
5$ which in turn is in principle able to move the multipoles away from
the direction of the dipole.

\section{Conclusions}
We have reanalysed the low $l$ multipole vectors of the three year WMAP
data. To this end, we presented a simple algorithm to compute these
multipole vectors which reduces this problem to finding the roots of a
polynomial. We argued for this algorithm by making contact with
Bloch coherent states on the sphere.

We pointed out that the alignment found earlier between the $l=2$ and
$l=3$ multipole vectors is really an anti-alignment with the dipole.
This anti-alignment is thus likely not cosmological as the dipole
itself is not cosmological but due to the sun's motion. Furthermore,
we found that this anti-alignment extends to three of the five
multipole vectors at $l=5$ and higher.

This anti-alignment with the dipole points towards a systematic error
stemming from the data processing. We could not conclusively pinpoint the
source but argued that a likely possibility is a non-linear mixing
the dipole and higher multipoles due to the running recalibration of
the WMAP detectors on the modulation of the dipole.  

It would be interesting to test this hypothesis with simulations using
the real data processing pipeline.

\acknowledgments It is a pleasure to acknowledge discussions
and useful correspondance with D. Schwarz, M. Tegmark, G. Weeks, and
C. Dennis.
We acknowledge the use of the Legacy Archive for Microwave Background
Data Analysis (LAMBDA).  Support for LAMBDA is provided by the NASA
Office of Space Science. This work was partly supported by the DFG
Schwerpunktprogramm 1096.

\appendix

\section{\label{olddata}Comparison with first year data}
The main text was based on the recently published temperature data of
the first three years of WMAP operation. The relevant foreground
cleaned map is the ``internal linear combination''.
\begin{table}
\caption{\label{multipoles}Multipole vectors up to $l=5$ using the
  first year Tegmark et al. map}
\begin{tabular}{|c||r|r|r|}
    \hline
  $l$&latitude&longitude&angle to dipole\\
  \hline\hline
  1&-96.15&48.25&0\\
  \hline
  2 & 11.26 & 16.64 & \bf 88.69\\
  & 118.87 & 25.13 & \bf 79.81\\
  \hline
  3 & 86.94 & 39.30 & \bf 87.59\\
  & 22.63 & 9.18 & \bf 78.61\\
  & -44.92 & 8.20 & 58.73\\
  \hline
  4&-26.49&26.86& 57.07\\
  &74.74&5.46& 54.31\\
  &-153.02&31.93& 45.30\\
  &-151.36&76.73& 35.58\\
  \hline
  5&44.67&33.54&\bf 88.97\\
  &172.84&3.07& \bf 88.39\\
  &98.70&38.50& \bf 87.75\\
  &-74.21&31.43& 23.63\\
  &-122.69&57.54&18.34\\
  \hline 
\end{tabular}
\end{table}

\begin{table}
\caption{\label{firstilc}Multipole vectors for first year ILC map data}
\begin{tabular}{|c||r|r|r|}
  \hline
  $l$&latitude&longitude&angle to dipole\\
  \hline\hline
2 & 15.55 & 3.21 & \bf 78.23\\
  & 120.95 & 19.79 & \bf 75.69\\
\hline
3 & 95.27 & 37.04 & \bf 85.89\\
  & 21.73 & 9.39 & \bf 79.31\\
  & -47.02 & 10.71 & 55.47\\
\hline
4 & 71.93 & 6.96 & 56.20\\
  & -28.31 & 30.23 & 53.66\\
  & -160.38 & 70.63 & 36.88\\
  & -142.64 & 39.46 & 34.10\\
\hline
5 & 100.79 & 38.52 & \bf 88.06\\
  & 176.05 & 1.20 & \bf 87.64\\
  & 40.12 & 37.00 & \bf 86.29\\
  & -73.53 & 34.23 & 21.90\\
  & -128.61 & 54.54 & 21.00\\
\hline
6 & 84.08 & 34.58 & \bf 82.83\\
  & 34.55 & 53.56 & 69.98\\
  & -23.00 & 17.06 & 66.21\\
  & -147.13 & 5.36 & 60.85\\
  & -74.51 & 20.96 & 32.34\\
  & -120.34 & 55.52 & 16.49\\
\hline
\end{tabular}
\end{table}

\begin{figure}
\caption{\label{1yilc}First year multipole vectors using the
  ILC map}
\includegraphics{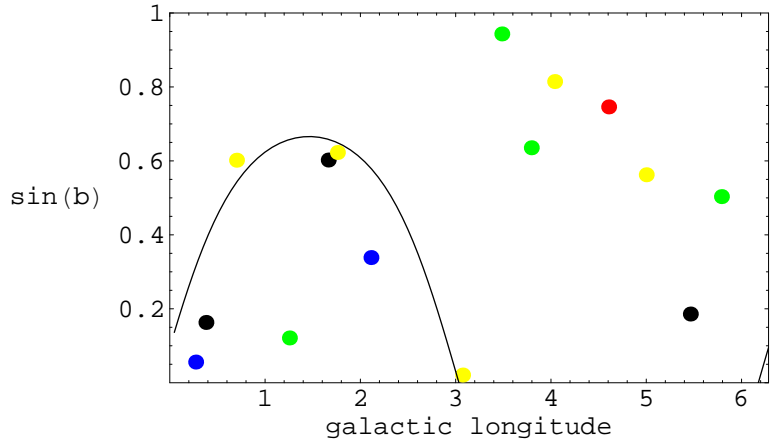}
\end{figure}

For comparison with the existing literature which is mainly based on
the quadratic Doppler shift corrected map of Tegmark et
al.\cite{Tegmark:2003ve}, cleaned map including the DQ higher order
Doppler correction\cite{Copi:2003kt}\footnote{We thank Max Tegmark and
  Greg Weeks for useful correspondence.}. Table \ref{multipoles} shows
the resulting multipole vectors while Figure \ref{year1tdqmap} shows them
in a plot of the northern hemisphere in galactic coordinates. There is
no equivalent to the Tegmark et al. map for the three year data. For
the reader to judge the influence of the different schemes for
foreground removal, we also list in table \ref{firstilc} the multipole
vectors for the ILC using the first year data only. The corresponding
plot is figure \ref{1yilc}


\bibliographystyle{apsrev} 
\bibliography{cmb}

\end{document}